\documentclass[conference]{IEEEtran}

\usepackage{multirow}
\usepackage{graphicx}
\usepackage{booktabs}
\usepackage{url}
\usepackage{cite}
\usepackage{amsmath}
\usepackage[final]{microtype}
\usepackage{algorithm}
\usepackage{algpseudocode}
\usepackage{amsthm}
\usepackage[table]{xcolor}
\usepackage{caption}
\usepackage[hidelinks]{hyperref}
\definecolor{stsoBlue}{RGB}{220, 235, 255}

\newcommand{\figref}[1]{\hyperref[#1]{Figure~\ref*{#1}}}
\newcommand{\tabref}[1]{\hyperref[#1]{Table~\ref*{#1}}}
\newcommand{\eqnref}[1]{\hyperref[#1]{Eq.~\ref*{#1}}}
\newcommand{\algoref}[1]{\hyperref[#1]{Algorithm~\ref*{#1}}}
\floatstyle{ruled}
\restylefloat{algorithm}
\algrenewcommand\algorithmicrequire{\textbf{Require:}}
\algrenewcommand\algorithmicensure{\textbf{Ensure:}}
\title{MoSE: Mode-Switching Expander for Mixed LLM Training and Inference\\

\author{
\IEEEauthorblockN{
Fan Yang\textsuperscript{1,2},
Ying Zhou\textsuperscript{3},
Binglei Wang\textsuperscript{1,2},
Zhenjie Zhou\textsuperscript{1,2},
Jialong Li\textsuperscript{2,*}
}
\IEEEauthorblockA{
\textsuperscript{1}Southern University of Science and Technology\\
\textsuperscript{2}Faculty of Computer Science and Artificial Intelligence, 
Shenzhen University of Advanced Technology\\
\textsuperscript{3}School of Electronic and Information Engineering, Beijing Jiaotong University
}
}
}

\begin{document}
\maketitle

\begin{abstract}
AI clusters increasingly run large language model (LLM) inference and training on the same fabric. Prefill-decode (P-D) disaggregation creates key-value (KV) cache transfers between prefill and decode groups, whereas training collectives and all-to-all traffic benefit from near-uniform global connectivity. A static sparse topology can therefore be poorly matched to one of the two traffic patterns. We present \emph{Mode-Switching Expander (MoSE)}, a reconfigurable expander that treats topology design as a fixed-degree edge-allocation problem. MoSE reallocates the same sparse edge budget toward direct P-D connectivity in inference-heavy modes and restores a uniform random regular expander in training-heavy modes. We evaluate MoSE using a 1024-group flow-level topology model, shortest-path routing, and two mixed workloads. Across 20 seeds, MoSE reduces average and 95th-percentile (P95) load-aware KV communication cost by 90.8\% and 91.9\% relative to Static-Training in the inference-heavy mode. In the training-heavy mode, it reduces average and P95 training communication cost by 22.7\% and 27.6\% relative to stale Static-Inference. These results show that coarse-grained topology switching can support both workload modes without additional ports or routing changes.
\end{abstract}


\section{Introduction}

AI clusters increasingly support both user-facing large language model (LLM) inference and model training on shared fabrics~\cite{sheng2023flexgen,zhong2024distserve,qin2024mooncake}. Such sharing can improve resource utilization because inference and training demands vary over time, allowing operators to reuse the same GPU and network resources across workload modes. At the network layer, however, this is not merely a scheduling problem. The interconnect must support latency-sensitive inference traffic during inference-heavy periods and high-volume collective traffic during training-heavy periods under the same sparse port budget.

This creates a fundamental tension: \emph{the topology that helps one workload mode can consume the edge budget needed by the other}. Prefill-decode (P-D) disaggregation creates concentrated demand across the P-D interface: prefill groups produce key-value (KV) cache state, which decode groups must receive before token generation proceeds~\cite{zhong2024distserve,qin2024mooncake,wang2025prefill,usami2026prefill,li2026revisiting,he2026efficient}. Training collectives, in contrast, often require broad global connectivity. All-to-all, all-reduce, and expert-parallel traffic can benefit from low diameter and near-uniform connectivity, making random regular expanders a natural fit~\cite{bernardi2026expanding}. Strengthening P-D connectivity can reduce KV-transfer pressure, but a static P-D bias sacrifices global expansion once the cluster shifts back to training-heavy operation. The core problem is therefore how to allocate the same fixed-degree edge budget between P-D connectivity and global expansion as workload modes change.

This observation motivates a mode-aware reconfigurable topology that reallocates, rather than expands, the fixed port budget. It must strengthen P-D connectivity when KV traffic dominates, restore global expansion when training dominates, and switch only at coarse workload-mode boundaries compatible with optical circuit switches~\cite{jouppi2023tpu,shou2025infinitehbd,li2025refly,liao2025mixnet,hu2026joint}. The resulting problem is to place the same sparse links where the active workload mode can use them most effectively.

We present \emph{Mode-Switching Expander (MoSE)}, which formulates reconfigurable expander design as a workload-mode-aware, fixed-degree edge-allocation problem. In inference-heavy mode, MoSE allocates more edges across the P-D cut. In training-heavy mode, it restores a uniform random regular expander for global communication. The key insight is that optical circuit switch (OCS) reconfiguration does not increase the port budget; it allows the same sparse optical fabric to realize different topology templates over time.

This paper makes the following contributions.
\begin{itemize}
  \setlength{\itemsep}{0.4em}
  \setlength{\parsep}{0pt}
  \setlength{\topsep}{0.2em}
  \setlength{\partopsep}{0pt}
  \setlength{\parskip}{0pt}

  \item We identify the fixed-degree edge-allocation tradeoff between P-D connectivity for disaggregated inference and global expansion for training communication.

  \item We introduce MoSE, reallocating the same sparse optical links between a P-D-biased topology for inference-heavy modes and a uniform expander for training-heavy modes.

  \item Through a 1024-group flow-level evaluation across 20 seeds, we show that MoSE reduces average and 95th-percentile (P95) load-aware KV communication cost by 90.8\% and 91.9\% relative to Static-Training in inference-heavy mode. In training-heavy mode, it reduces average and P95 training communication cost by 22.7\% and 27.6\% relative to stale Static-Inference.
\end{itemize}
\vspace{-0.85em}

\section{Background and Motivation}

Mixed LLM clusters no longer exhibit a single traffic geometry. The same optical fabric may carry latency-sensitive inference traffic in one period and high-volume training traffic in another. Under a fixed port budget, topology design therefore becomes an edge-allocation problem: how should the same sparse links be distributed across different traffic cuts?

P-D disaggregated inference concentrates traffic across the prefill-decode partition~\cite{zhong2024distserve,qin2024mooncake,wang2025prefill,li2026revisiting}. Prefill groups produce KV cache, which decode groups consume before token generation proceeds. When the two phases run on separate GPU groups, the resulting traffic has a bipartite-like structure between prefill and decode groups. Increasing direct P-D connectivity can therefore reduce KV-transfer hops and alleviate bottleneck.

Training communication has a different preference. Large-scale training combines global collectives, such as all-reduce and all-to-all, with structured communication across parallel groups~\cite{li2025refly,liao2025mixnet}. We use directed all-to-all traffic as a representative stress case because it exercises global connectivity across the fabric. A uniform sparse expander fits this pattern by distributing edges broadly and keeping paths short~\cite{bernardi2026expanding}. Under a fixed degree, however, increasing P-D cut capacity necessarily reduces the edge budget available for uniform global expansion.

This mismatch suggests that topology should change with workload mode rather than with individual requests or packets. An OCS fabric can strengthen P-D connectivity when KV cache traffic dominates and restore a uniform expander when training traffic dominates~\cite{jouppi2023tpu,shou2025infinitehbd,li2025refly,liao2025mixnet,hu2026joint}. MoSE realizes this coarse-grained adaptation by reallocating the same fixed-degree edge budget across workload modes.

\section{Model and Design}

We model the cluster as a group-level graph. A node represents a rack, GPU island, or similar resource group, and an undirected edge represents one bidirectional inter-group optical circuit or link bundle. For topology-level accounting, traffic in both directions is aggregated on the same normalized unit-capacity edge. The graph is simple, connected, and $d$-regular. With $N=1024$ groups and $d=4$, every topology contains 2048 inter-group optical circuits.

A flow is denoted by $f=(s_f,t_f,x_f,k_f)$, where $s_f$ and $t_f$ are its source and destination, $x_f$ is its demand, and $k_f\in\{\mathrm{kv},\mathrm{tr}\}$ is its traffic class. Each flow is routed over one shortest path. We report maximum and P95 link load, average and P95 load-aware communication cost, hop count, and KV path structure.

Workload changes alter node roles rather than cluster size. The inference-heavy workload contains 410 prefill, 410 decode, and 204 training groups, whereas the training-heavy workload contains 102 prefill, 102 decode, and 820 training groups. Within each workload mode, all policies use the same node partition and traffic matrix and differ only in edge placement.

\begin{figure*}[t]
  \centering
  \includegraphics[width=\textwidth]{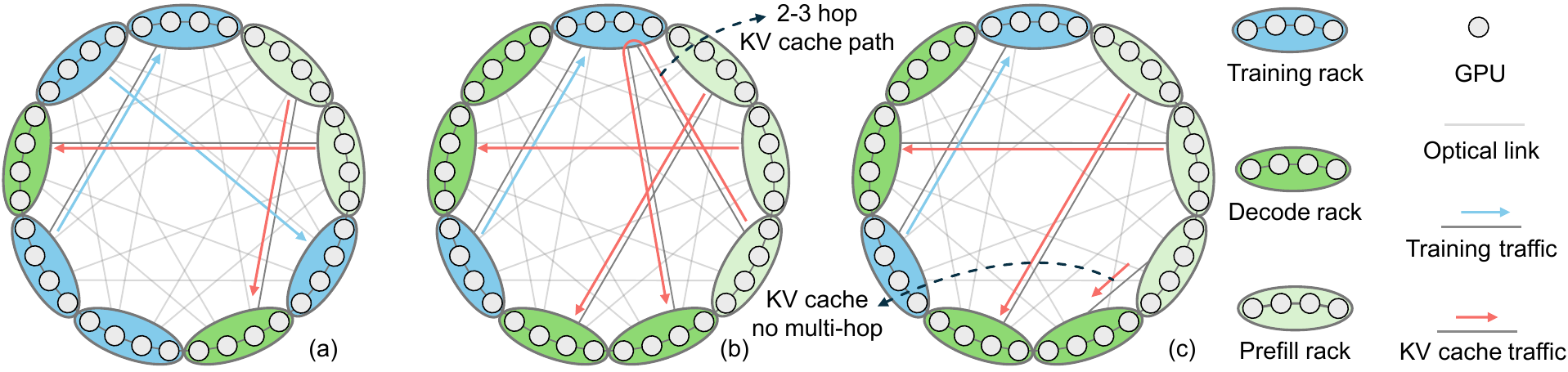}
  \caption{MoSE reconfiguration. (a) Training and inference share one optical fabric. (b) As inference demand increases, KV cache transfers may traverse multiple hops. (c) OCS reconfiguration installs direct P-D links for one-hop KV transfer.}
  \label{fig:mode-aware-schematic}
\end{figure*}

\algoref{alg:mose} summarizes the topology construction. MoSE keeps the vertex set and per-node degree fixed and selects a topology template according to the workload mode. In inference-heavy mode, it reserves $r_{\mathrm{pd}}$ ports at each prefill and decode group for an $r_{\mathrm{pd}}$-regular bipartite subgraph and randomly fills the remaining degree budget while preserving graph simplicity and connectivity. In training-heavy mode, it instantiates a uniform random regular expander. We use $r_{\mathrm{pd}}=d-1=3$ for the P-D-biased topology, retaining one residual port per P-D group for global connectivity. The compromise topology uses $r_{\mathrm{pd}}=1$. Thus, MoSE changes only edge placement, not routing, scheduling, or port count.

\begin{algorithm}[t]
\small
\caption{MoSE topology construction}
\label{alg:mose}
\begin{algorithmic}[1]
\Require groups $\mathcal{G}$, degree $d$, workload mode $m$, roles $(\mathcal{P},\mathcal{D},\mathcal{T})$, P-D degree $r_{\mathrm{pd}}$
\Ensure simple connected undirected $d$-regular topology $H_m=(V,E_m)$
\State $V \gets \mathcal{G}$
\If{$m=\mathsf{InferenceHeavy}$}
  \State $E_m \gets \emptyset$
  \State $b(v)\gets d$ for each $v\in V$
  \State $E_{\mathrm{pd}}\gets\Call{BuildPDBipartite}{\mathcal{P},\mathcal{D},r_{\mathrm{pd}}}$
  \State $E_m\gets E_m\cup E_{\mathrm{pd}}$
  \State update residual degree budgets $b(\cdot)$
  \State $E_{\mathrm{fill}}\gets\Call{RandomFill}{V,E_m,b}$
  \State $E_m\gets E_m\cup E_{\mathrm{fill}}$
\Else
  \State $E_m\gets\Call{RandomRegularEdges}{V,d}$
\EndIf
\State repair $E_m$ until $(V,E_m)$ is simple and connected
\State assert $\deg(v)=d$ for all $v\in V$
\State \Return $H_m=(V,E_m)$
\end{algorithmic}
\end{algorithm}

\subsection{Topology, Traffic, and Metrics}

We write a topology as $H=(V,E)$ and compare simple connected $d$-regular graphs on the same vertex set. By the handshaking identity, every evaluated topology has the same number of inter-group links:
\begin{equation}
  |E|=\frac{Nd}{2}.
  \label{eq:edge-budget}
\end{equation}
\eqnref{eq:edge-budget} is the accounting constraint: reconfiguration can relocate edges but cannot create additional links.

We consider three topology templates. The uniform topology $H^{\mathrm{U}}$ is a role-agnostic random regular graph. The P-D-biased topology $H^{\mathrm{PD}}$ reserves $d-1$ degrees at each prefill and decode group for P-D connectivity. The compromise topology $H^{\mathrm{C}}$ reserves one degree for P-D connectivity and uses the remaining budget for global expansion. The P-D-biased and compromise templates are instantiated for the inference-heavy P-D assignment and remain fixed for the corresponding static baselines.

\textbf{Static-Training} always uses $H^{\mathrm{U}}$. \textbf{Static-Inference} always uses $H^{\mathrm{PD}}$, including when its P-D structure becomes stale in training-heavy mode. \textbf{Static-Compromise} always uses $H^{\mathrm{C}}$. \textbf{MoSE} uses $H^{\mathrm{PD}}$ in inference-heavy mode and switches to $H^{\mathrm{U}}$ in training-heavy mode.

For reporting direct P-D connectivity, we define
\begin{equation}
  E_{\mathrm{PD}}(H)=
  \bigl\{\{u,v\}\in E:
  u\in\mathcal{P},\,v\in\mathcal{D}
  \ \text{or}\
  u\in\mathcal{D},\,v\in\mathcal{P}\bigr\}.
  \label{eq:pd-edges}
\end{equation}
We report $|E_{\mathrm{PD}}(H)|$ as the number of direct P-D edges. This quantity is distinct from the graph-theoretic cut $\delta_H(\mathcal{P})$, which also contains edges between prefill and training groups.

Each workload mode $m$ induces KV cache flows $F^{\mathrm{kv}}_m$ and training flows $F^{\mathrm{tr}}_m$. Let $F_m=F^{\mathrm{kv}}_m\cup F^{\mathrm{tr}}_m$. In our experiments, $|F^{\mathrm{kv}}_m|=\rho^{\mathrm{kv}}_m|F_m|$, where $\rho^{\mathrm{kv}}_m$ is 0.70 in inference-heavy mode and 0.30 in training-heavy mode. KV cache flows follow the P-D assignment of the active workload mode, while training flows sample source-destination pairs uniformly among training groups.

For each flow $f$, routing selects one shortest path $p_H(f)$:
\begin{equation}
  p_H(f)\in\arg\min_{p:s_f\rightarrow t_f}|p|.
  \label{eq:shortest-path}
\end{equation}
The simulator computes shortest paths using breadth-first search. Ties among equal-length shortest paths are broken uniformly at random using the same rule for all policies.

The load contributed to edge $e$ by traffic class $k$ is
\begin{equation}
  \ell^k_e(H,F_m)
  =
  \sum_{\substack{f\in F^k_m\\e\in p_H(f)}}x_f.
  \label{eq:class-edge-load}
\end{equation}
The total load on edge $e$ is $\ell_e(H,F_m)=\sum_k\ell^k_e(H,F_m)$. For traffic class $k$, maximum link load is
\begin{equation}
  L^k_m(H)=\max_{e\in E}\ell^k_e(H,F_m),
  \label{eq:class-maxload}
\end{equation}
and the overall maximum link load is $L_m(H)=\max_{e\in E}\ell_e(H,F_m)$. We define link-load P95 analogously over the complete edge-load vector.

We define the load-aware communication cost of flow $f$ as the accumulated mixed-traffic load along its selected path:
\begin{equation}
  C_f(H,F_m)=
  \sum_{e\in p_H(f)}\ell_e(H,F_m).
  \label{eq:flow-comm-cost}
\end{equation}
This metric captures the contention encountered along a path and serves as a topology-level proxy rather than wall-clock communication time. For traffic class $k$, its average communication cost is
\begin{equation}
  \bar{C}^k_m(H)=
  \operatorname{avg}_{f\in F^k_m}C_f(H,F_m),
  \label{eq:avg-comm-cost}
\end{equation}
and its tail communication cost is
\begin{equation}
  C^{k,95}_m(H)=
  \operatorname{P95}_{f\in F^k_m}C_f(H,F_m).
  \label{eq:p95-comm-cost}
\end{equation}
Figures use $L^k_m$, $\bar{C}^k_m$, and $C^{k,95}_m$ for the traffic class being analyzed.

Let $h_H(f)=|p_H(f)|$ denote hop count. We additionally report the direct KV ratio
$R_{\mathrm{direct}}(H,m)=
\Pr_{f\in F^{\mathrm{kv}}_m}[h_H(f)=1]$
and the long-path KV ratio
$R_{3+}(H,m)=
\Pr_{f\in F^{\mathrm{kv}}_m}[h_H(f)\geq3]$.

The fixed edge budget also yields a generic cut-pressure lower bound. For $S\subset V$, let
\[
  \delta_H(S)=
  \{\{u,v\}\in E:u\in S,\,v\in V\setminus S\}
\]
and let the bidirectional demand crossing the cut be
\[
  D_m(S)=
  \sum_{\substack{f\in F_m\\
  |\{s_f,t_f\}\cap S|=1}}x_f.
\]
Any routing with maximum link load $L_m(H)$ must carry this demand over the cut edges:
\begin{equation}
  L_m(H)\geq
  \frac{D_m(S)}{|\delta_H(S)|}.
  \label{eq:cut-pressure}
\end{equation}
The bound exposes the structural tradeoff under a fixed degree budget. P-D-biased templates devote more edges to direct P-D connectivity, whereas uniform expanders distribute the same edge budget broadly to support global training traffic.

\subsection{Optical Fabric Mapping}

MoSE maps naturally to an optical circuit-switched fabric, although our evaluation remains topology-level. A group with degree $d$ requires $d$ OCS-facing ports, so $N=1024$ and $d=4$ correspond to 4096 optical endpoints and 2048 active bidirectional circuits. These circuits may be realized by a large-radix OCS fabric or multiple OCS planes, with one group-facing port on each plane. This abstraction is consistent with reconfigurable AI fabrics that change physical connectivity at a coarse control timescale rather than at packet granularity~\cite{jouppi2023tpu,shou2025infinitehbd,li2025refly,liao2025mixnet,hu2026joint}.

Mode switches do not change the per-group port budget; MoSE only changes how the optical endpoints are paired. Each circuit may operate at the attached transceiver rate, such as 400G or 800G, while the graph model normalizes its capacity. We assume that circuits affected by a mode switch are reconfigured after their active transfers drain. Hitless or staged reconfiguration across multiple OCS planes is outside the scope of this topology-level study.

We do not model transient reconfiguration delay, insertion loss, transport dynamics, or packet-level flow completion time. These factors determine absolute transfer time and deployment overhead, whereas our evaluation isolates topology-level connectivity and contention under a fixed port budget.

In the following, S-Train, S-Inf, and S-Comp denote Static-Training, Static-Inference, and Static-Compromise, respectively.

\begin{figure*}[t]
  \centering
  \begin{minipage}[t]{0.48\textwidth}
    \centering
    \includegraphics[width=\linewidth]{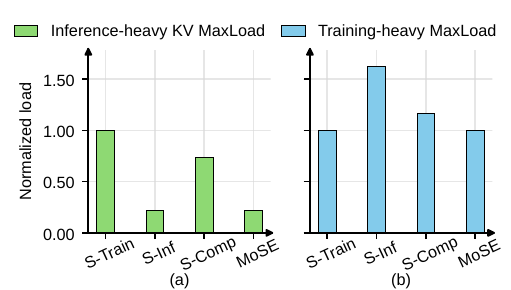}
    \caption{Bottleneck load. (a) KV cache MaxLoad under inference-heavy demand. (b) Training MaxLoad under training-heavy demand.}
    \label{fig:maxload-summary}
    \vspace{0.6em}

    \includegraphics[width=\linewidth]{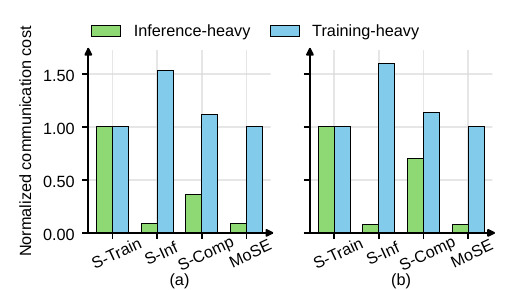}
    \caption{Load-aware KV cache communication cost. (a) Average KV cache communication cost. (b) P95 KV cache communication cost.}
    \label{fig:kv-cache-time}
  \end{minipage}\hfill
  \begin{minipage}[t]{0.48\textwidth}
    \centering
    \includegraphics[width=\linewidth]{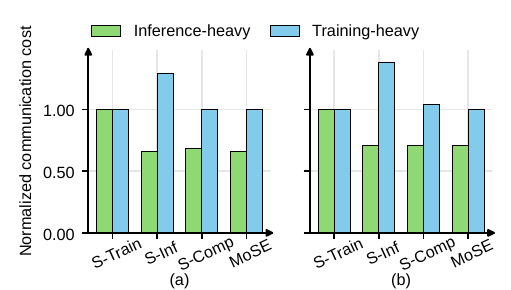}
    \caption{Load-aware training communication cost. (a) Average training communication cost. (b) P95 training communication cost.}
    \label{fig:training-time}
    \vspace{0.6em}

    \includegraphics[width=\linewidth]{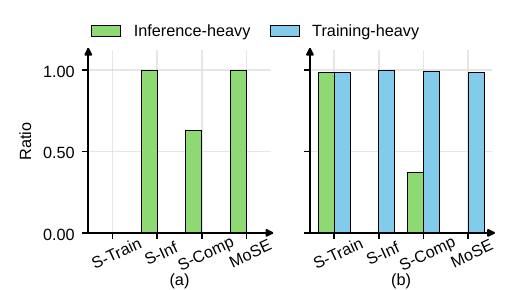}
    \caption{KV cache path structure. (a) Direct KV cache ratio. (b) 3+ hop KV cache ratio.}
    \label{fig:kv-cache-path}
  \end{minipage}
\end{figure*}

\section{Evaluation}

We evaluate the four policies with $N=1024$ groups and degree $d=4$. Each result is averaged over 20 random seeds covering topology generation and traffic sampling. The inference-heavy workload contains 410 prefill groups, 410 decode groups, 204 training groups, and $100{,}000$ unit-size flows, with 70\% KV cache traffic and 30\% training traffic. The training-heavy workload contains 102 prefill groups, 102 decode groups, 820 training groups, and the same number of flows, with 30\% KV cache traffic and 70\% training traffic. Within each workload mode, the P-D assignment is assumed to remain stable over one topology reconfiguration interval. KV cache traffic follows this assignment, while training traffic samples T-T pairs uniformly. For each seed, all policies use the same node partition and traffic matrix within each workload mode, so the comparison isolates the effect of edge placement.

We report three groups of metrics. Maximum and P95 link load capture network bottlenecks. Average and P95 load-aware communication cost capture the contention accumulated along KV and training paths. Direct-KV and 3+-hop ratios describe the resulting KV path structure. The communication-cost metric is a topology-level proxy rather than wall-clock transfer time. \tabref{tab:mose-results} reports the averaged results. Table headers abbreviate average hop count and direct KV cache ratio as AvgHop and DirectKV, respectively.

\begin{table*}[t]
  \centering
  \caption{Flow-Level Results Under Different Workloads}
  \label{tab:mose-results}
  \setlength{\tabcolsep}{3.0pt}
  \renewcommand{\arraystretch}{1.25} 
  \begin{tabular}{@{}ll*{8}{c}@{}}
    \toprule
    {\textbf{Workload}} & {\textbf{Method}} & {\textbf{P-D Cut}} & {\textbf{MaxLoad $\downarrow$}} & {\textbf{P95Load $\downarrow$}} & {\textbf{LoadImb. $\downarrow$}} & {\textbf{AvgHop $\downarrow$}} & {\textbf{KV AvgHop $\downarrow$}} & {\textbf{DirectKV $\uparrow$}} & {\textbf{KV 3-hop+ $\downarrow$}} \\
    \midrule
    
    & Static-Training & 658.45 & 1041.95 & 596.04 & 0.809 & 5.66 & 5.66 & 0.004 & 0.983 \\
    & Static-Inference & 1331.65 & \textbf{529.35} & \textbf{287.80} & \textbf{0.636} & \textbf{2.46} & \textbf{1.00} & \textbf{1.000} & \textbf{0.000} \\
    & Static-Compromise & 925.95 & 804.15 & 412.43 & 0.719 & 3.62 & 2.75 & 0.626 & 0.369 \\
    \rowcolor{stsoBlue} 
    \cellcolor{white}\multirow{-4}{*}{\textbf{Inference-heavy}} 
    & \textbf{MoSE} & 1331.65 & \textbf{529.35} & \textbf{287.80} & \textbf{0.636} & \textbf{2.46} & \textbf{1.00} & \textbf{1.000} & \textbf{0.000} \\
    \midrule
    
    & Static-Training & 39.95 & \textbf{1157.75} & \textbf{534.46} & \textbf{0.547} & \textbf{5.65} & \textbf{5.64} & \textbf{0.005} & \textbf{0.983} \\
    & Static-Inference & 6.60 & 1885.60 & 736.72 & 0.693 & 6.27 & 6.38 & 0.000 & 0.997 \\
    & Static-Compromise & 14.00 & 1347.15 & 563.59 & 0.600 & 5.67 & 5.70 & 0.001 & 0.993 \\
    \rowcolor{stsoBlue} 
    \cellcolor{white}\multirow{-4}{*}{\textbf{Training-heavy}} 
    & \textbf{MoSE} & 39.95 & \textbf{1157.75} & \textbf{534.46} & \textbf{0.547} & \textbf{5.65} & \textbf{5.64} & \textbf{0.005} & \textbf{0.983} \\
    
    \bottomrule
  \end{tabular}
\end{table*}

\begin{table}[t]
  \centering
  \caption{Path-Cost Values for Headline Results}
  \label{tab:path-cost-headline}
  \small
  \setlength{\tabcolsep}{0pt}
  \renewcommand{\arraystretch}{1.08}
  \begin{tabular*}{\columnwidth}{@{\extracolsep{\fill}}llccc@{}}
    \toprule
    \textbf{Metric} & \textbf{Base} & \textbf{Base Cost} & \textbf{MoSE} & \textbf{Red.} \\
    \midrule
    Inf. KV Avg & S-Train & 2419.68 & 222.16 & 90.8\% \\
    Inf. KV P95 & S-Train & 3592.60 & 290.41 & 91.9\% \\
    Training Avg & S-Inf & 2039.17 & 1575.36 & 22.7\% \\
    Training P95 & S-Inf & 3475.41 & 2517.51 & 27.6\% \\
    \bottomrule
  \end{tabular*}
\end{table}

\textbf{Bottleneck load.}
\figref{fig:maxload-summary}(a) reports KV-specific MaxLoad under the inference-heavy workload. MoSE matches Static-Inference because both use the same P-D-biased topology in this mode. The aggregate mixed-traffic results in \tabref{tab:mose-results} show the same trend: relative to Static-Training, MoSE reduces overall MaxLoad from 1041.95 to 529.35 and P95 link load from 596.04 to 287.80. Static-Compromise lies between the two static endpoints because it allocates only part of the degree budget to direct P-D connectivity. \figref{fig:maxload-summary}(b) reports training-specific MaxLoad under the training-heavy workload. At the aggregate level, retaining the stale P-D-biased topology increases MaxLoad from 1157.75 under Static-Training to 1885.60 under Static-Inference. MoSE avoids this increase by returning to the uniform expander.

\textbf{Load-aware communication cost.}
Maximum link load captures the most congested edge, but it does not describe the accumulated contention along a routed flow. \figref{fig:kv-cache-time} reports average and P95 load-aware KV communication cost, while \tabref{tab:path-cost-headline} gives the corresponding raw values. Under inference-heavy demand, MoSE reduces average and P95 KV communication cost by 90.8\% and 91.9\% relative to Static-Training. \figref{fig:training-time} shows the complementary result for training traffic. Under training-heavy demand, MoSE reduces average and P95 training communication cost by 22.7\% and 27.6\% relative to stale Static-Inference, while matching Static-Training. MoSE therefore reaches the corresponding mode-specific static topology in both evaluated workloads: the P-D-biased topology under inference-heavy demand and the uniform expander under training-heavy demand.

\textbf{Path mechanism.}
\figref{fig:kv-cache-path} helps explain the KV communication-cost reduction through changes in path structure. \figref{fig:kv-cache-path}(a) reports the Direct-KV ratio, which increases from 0.004 under Static-Training to 1.000 under MoSE in the inference-heavy workload. \figref{fig:kv-cache-path}(b) reports the 3+-hop KV ratio, which decreases from 0.983 to 0.000 under the same workload. Static-Compromise again occupies an intermediate point, with a Direct-KV ratio of 0.626. In the training-heavy workload, direct P-D ratios remain near zero because MoSE uses a uniform expander, while the static P-D-biased templates retain the inference-heavy P-D assignment. These results suggest that reallocating edges toward the active P-D assignment can shorten KV paths when inference traffic dominates.

\textbf{Scope of the evaluation.}
The evaluation isolates a topology-level question: under a fixed-degree optical-link budget, should sparse links remain broadly distributed for training traffic, or should more of them be allocated to P-D connectivity when KV traffic dominates? The flow-level model separates this question from transport control, GPU scheduling, and serving policy. It treats each graph node as a rack-scale group and uses link load and load-aware communication cost to characterize topology pressure. Factors such as KV cache size, prompt length, model size, data type, collective implementation, ECMP routing, OCS reconfiguration cost, and the host-ToR-OCS hierarchy may affect absolute communication performance and the magnitude of the observed gains. Our evaluation instead focuses on the underlying edge-allocation tradeoff between concentrated P-D connectivity and broad global expansion.

\section{Related Work}

\textbf{LLM serving and inference communication.}
LLM serving systems optimize memory and execution placement,
prefill-decode separation, and KV-cache management to improve
inference throughput and goodput~\cite{sheng2023flexgen,zhong2024distserve,qin2024mooncake}.
Follow-up work studies P-D aggregation, accelerator-specific
behavior, energy efficiency, and repeated KV transfers in
multi-round inference~\cite{wang2025prefill,usami2026prefill,li2026revisiting,he2026efficient}.
Communication-aware systems further coordinate model deployment
and communication scheduling for MoE inference~\cite{li2025optimizing}.
MoSE is complementary: it keeps serving and placement policies
fixed and adapts the inter-group optical topology to the resulting
P-D traffic.

\textbf{AI-cluster communication and network design.}
Training collectives and MoE all-to-all communication require
balanced global connectivity, motivating communication-aware
load balancing~\cite{xu2026rails}. Recent work also explores
hyperscale network architectures designed specifically for AI
systems~\cite{cai2026matryoshka}. MoSE focuses on a different
dimension: temporal workload shifts within one shared fabric,
reallocating a fixed-degree link budget between inference-oriented
and training-oriented topologies.

\textbf{Expanders and reconfigurable optical fabrics.}
Random-graph data center networks provide sparse global
connectivity, while optical circuit switching enables physical
topologies to change over time~\cite{bernardi2026expanding,jouppi2023tpu,shou2025infinitehbd,li2025refly,liao2025mixnet,hu2026joint}.
Prior work has studied topology selection in reconfigurable
networks~\cite{yu2025topofair}, as well as routing over
fast-switched optical fabrics and for latency-sensitive
flows~\cite{li2025unlocking,li2022hop}. MoSE combines
mode-aware topology selection with expander design: rather
than optimizing routes, it fixes shortest-path routing and moves
the same sparse edge budget between the P-D cut and uniform
global expansion.

\section{Conclusion}

We presented MoSE, a mode-switching expander for mixed LLM training and inference. MoSE reallocates a fixed-degree optical edge budget between direct P-D connectivity for KV cache transfer and uniform expansion for training traffic. In a 1024-group flow-level evaluation, it reduces load-aware KV communication cost in inference-heavy mode while avoiding the higher training cost of a stale P-D-biased topology. These results suggest that coarse-grained topology reconfiguration can adapt a shared optical fabric to changing LLM workload modes without adding ports or modifying routing.

\bibliographystyle{IEEEtran}
\bibliography{reference}

\end{document}